\documentclass[conference,a4paper]{IEEEtran}
\IEEEoverridecommandlockouts

\IEEEsettopmargin{t}{30mm}
\IEEEquantizetextheight{c}
\IEEEsettextwidth{14mm}{14mm}
\IEEEsetsidemargin{c}{0mm}

\usepackage[utf8]{inputenc}
\usepackage[T1]{fontenc}
\usepackage{amsmath,amsthm,amsfonts,amssymb}
\usepackage{graphicx}
\usepackage{color,url}

\usepackage{datetime}

\usepackage{nicefrac}

\usepackage{enumitem} % advanced enumerations

\usepackage[english]{babel}

\usepackage{verbatim} % just to use {comment} environment

\usepackage[algoruled,vlined,linesnumbered,nofillcomment]{algorithm2e} % allows vertical lines for blocks

\usepackage[capitalize]{cleveref} % "clever" references with \cref{} (no "Theorem~" needed)
  \crefname{equation}{}{}
  \crefname{thm}{Theorem}{Theorems}
  \crefname{lemma}{Lemma}{Lemmas}
  \crefname{cor}{Corollary}{Corollaries}
  \crefname{define}{Definition}{Definitions}
  \crefname{note}{Note}{Notes}
  \crefname{propose}{Proposition}{Propositions}
  \crefname{appsec}{Appendix}{Appendices}

\usepackage{booktabs}  % improved lines in table
\usepackage{tabularx}  % for 'tabularx' environment and 'X' column type
\usepackage{multirow}

\usepackage{bm}
\usepackage{balance}

\theoremstyle{plain} % default
\newtheorem{thm}{Theorem}
\newtheorem{lemma}{Lemma}
\newtheorem{cor}{Corollary} % numbering is needed, as we refer some of corrolaries

\theoremstyle{definition}
\newtheorem{define}{Definition}

\newcommand{\Prob}[2][]{\mathbb P_{#1} \left\{ #2 \right\}}

\newcommand{\transp}{^{\intercal}}
\DeclareMathOperator{\rank}{rank}

\DeclareMathOperator{\diag}{diag}

\renewcommand{\vec}[1]{\bm{#1}}
\newcommand{\mtrx}[1]{\mathsf{#1}}
\newcommand{\const}[1]{\textnormal{\usefont{U}{eur}{m}{n}\selectfont #1}} % Euler font for special constants

\DeclareMathOperator{\HH}{H}
\DeclareMathOperator{\II}{I}
\DeclareMathOperator{\dlog}{dlog}

\newcommand{\Ring}{\mathcal{R}}
\newcommand{\GF}{\mathbb{F}}
\newcommand{\Ints}{\mathbb{Z}}

\newcommand{\unif}{\mathfrak{U}}
\newcommand{\toZm}{^{\langle m \rangle}}
\newcommand{\hh}{h}

\usepackage[normalem]{ulem} % strikethrough font

\begin{document}
	
\title{On the Capacity of Private Monomial Computation}

\author{\IEEEauthorblockN{Yauhen Yakimenka, Hsuan-Yin Lin, and Eirik Rosnes}\\%[1ex]
	\vspace{-4mm}\IEEEauthorblockA{Simula UiB, N–5008 Bergen, Norway\\
		Email: \{yauhen, lin, eirikrosnes\}@simula.no}
	%\thanks{The work of \dots}
}%

\maketitle

\begin{abstract}
In this work, we consider private monomial computation (PMC) for replicated noncolluding databases. In PMC, a user wishes to privately retrieve an arbitrary multivariate monomial from a candidate set of monomials in $f$ messages over a finite field $\GF_q$, where $q=p^k$ is a power of a prime $p$ and $k \ge 1$, replicated over $n$ databases. We derive the PMC capacity under a technical condition on $p$ and for asymptotically large $q$. The condition on $p$ is satisfied, e.g.,  for large enough $p$. Also, we present a novel PMC scheme for arbitrary $q$ that is capacity-achieving in the asymptotic case above. Moreover, we present formulas for the entropy of a multivariate monomial and for a set of monomials in uniformly distributed random variables over a finite field, which are used in the derivation of the capacity expression.
\end{abstract}

\section{Introduction}
The concept of private computation (PC) was introduced independently by Sun and Jafar \cite{SunJafar19_2} and Mirmohseni and Maddah-Ali \cite{MirmohseniMaddahAli18_1}. In PC, a user wishes to compute a function of the messages stored in a set of databases without revealing any information about the function to any of the databases. PC can be seen as a generalization of private information retrieval (PIR). In PIR, a user wants to retrieve a single message from the set of databases privately. Applications of PC include, in principle, all scenarios where insights about certain actions of the user should be kept private. One practical motivation for considering arbitrary functions is that of \emph{algorithmic privacy}, as protecting the identity of an algorithm running in the cloud could be even more critical than data privacy in some scenarios. Not only could the algorithm be valuable, but also in some cases, parameters of the algorithm carry lifetime secrets such as biological information of individuals \cite{MirmohseniMaddahAli18_1}.

The capacity in the linear case, i.e., the computation of arbitrary linear combinations of the stored messages, has been settled for both replicated \cite{SunJafar19_2} and coded \cite{ObeadKliewer18_1,ObeadLinRosnesKliewer18_1} databases. In the coded databases scenario, the messages are encoded by a linear code before being distributed and stored in a set of databases. Interestingly, the capacity in the linear case is equal to the corresponding PIR capacity for both replicated and coded databases. The monomial case was recently considered in \cite{ObeadLinRosnesKliewer19_1,ObeadLinRosnesKliewer19_2}. However, the presented achievable schemes have a PC rate, defined here as the ratio between the \emph{smallest} desired amount of information and the total amount of downloaded information, that in general is strictly lower than the best known converse bound for a finite number of messages. PC schemes in the coded case for arbitrary polynomials were considered by Karpuk and Raviv in \cite{Karpuk18_1,RavivKarpuk19_1}, and recently improved in \cite{ObeadLinRosnesKliewer19_1} when the number of messages is small. The capacity of private polynomial computation for coded databases remains open.

In this work, we first derive formulas for the entropy of a multivariate monomial and a set of monomials in uniformly distributed random variables over a finite field. We then present a novel PC scheme for multivariate monomials in the messages stored in a set of replicated noncolluding databases. The key ingredient of the scheme is the use of discrete logarithms. The discrete logarithm in the multiplicative group of a finite field of order $q=p^k$ ($p$ is a prime and $k \geq 1$) is a bijection to the integer ring of size $q-1$, mapping multiplication to addition. Hence, the discrete logarithm maps multivariate monomial retrieval to linear function retrieval, given that none of the messages is the zero element. The latter holds with probability approaching one as $q$ becomes large. 
The corresponding PC rate in this limiting case is derived using the entropy formulas from the first part of the paper. When the candidate set of multivariate monomials is fixed (i.e., independent of $q$), the PC rate converges to the PIR capacity for any number of messages stored in the databases, under a technical condition on $p$ and  as $q$ goes to infinity. The condition on $p$ is satisfied, e.g., for large enough $p$. Also, the presented monomial computation scheme is capacity-achieving in this asymptotic case.

\section{Preliminaries}

\subsection{General Definitions and Notation}

Throughout the paper, vectors are denoted by bold font and matrices are written as sans-serif capitals.

We work with different algebraic structures: the ring of integers $\Ints$, rings of residuals $\Ints_m$ for integers $m > 1$, and finite fields $\GF_q$,  where $q=p^k$ is a power of a prime $p$ and $k \ge 1$. %\footnote{Further in the paper, we always imply $q = p^k$, for prime $p$ and $k \ge 1$.} 
Occasionally, $\mathcal R$ denotes any of these structures. We often use the connection between $\Ints$ and $\Ints_m$. In principle, any element in $\Ints$ can be considered as an element of $\Ints_m$, with correspondence of addition and multiplication. If an expression consists of both integers and elements of $\Ints_m$, we assume all operations are over $\Ints_m$. When we need to stress that an element is in $\Ints_m$, we write $a\toZm\in\Ints_m$ for $a \in \Ints$.  The same notation is used for matrices, e.g., $\mtrx A\toZm$ has entries $a\toZm_{ij}\in\Ints_m$ for $a_{ij}\in\Ints$.

Any $a \in \Ints$ can be viewed as $a^{\langle p \rangle} \in \Ints_p = \GF_p \subseteq \GF_q$. Operations on such elements of $\GF_q$ are modulo $p$, as $p$ is the \emph{characteristic} of $\GF_q$, i.e., the minimum positive integer $l$ such that $l \cdot \alpha = 0$ for all $\alpha \in \GF_q$. Analogously, $\mtrx A \in \Ints^{s \times t}$ can be viewed as $\mtrx A^{\langle p \rangle} \in \GF_q^{s \times t}$. Note the difference between $\mtrx A^{\langle p \rangle} \in \GF_q^{s \times t}$ and $\mtrx A^{\langle q \rangle} \in \Ints_q^{s \times t}$ for $q=p^k$ and $k > 1$.

%Sometimes, to make a distinction for variables that relate to $\Ints_m$ as opposed to $\Ints$ and $\GF_q$ we use the prime symbol, e.g., $r'$ and $\bm L'$.
The \emph{multiplicative group} $\GF_q^* = \GF_q \setminus \{0\}$ is cyclic (cf. \cite[Thm.~2.18]{basicalgebraI}), and it is possible to define a \emph{discrete logarithm} function\footnote{Strictly speaking, $\dlog$ requires fixing a particular generator of $\GF_q^*$.
% We assume that the implied generator is the same throughout the paper.
} $\dlog : \GF_q^* \to \Ints_{q-1}$, which is an isomorphism between $(\GF_q^*, \times)$ and $(\Ints_{q-1}, +)$.

%"Sometimes when we need to make a distinction Z_m as opposed to Z and F_q, we use the prime symbol (')". Or something like that

We write $[a]\triangleq\{1,\ldots,a\}$ for a positive integer $a$. The \emph{greatest common divisor (gcd)} of $a_1, \dotsc, a_s \in \Ints$ is denoted by $\gcd(a_1, \dotsc, a_s)$, with the convention $\gcd(0,\dotsc,0) \triangleq 0$ and $\gcd(a_1\toZm, \dotsc, a_s\toZm, m) \triangleq \gcd(a_1, \dotsc, a_s, m)$. We write $a \mid b$ when $a$ divides $b$, and $a \nmid b$ otherwise. The binomial coefficient of $a$ over $b$ (both nonnegative integers) is denoted by $\binom{a}{b}$  where $\binom{a}{b}=0$ if $a <b$. The transpose of $\mtrx A$ is denoted by $\mtrx A\transp$.

%
%Consider a matrix $\mtrx A \in \mathcal R^{s \times t}$ and an integer $k$. Remove  $s-k$ arbitrary rows and $t-k$ arbitrary columns from $\mtrx A$. We call the 
%
A $k \times k$ \emph{minor} in $\mathcal R$ of a matrix $\mtrx A \in \mathcal R^{s \times t}$, for  a positive integer $k$, is the determinant of a $k \times k$ submatrix of $\mtrx A$ obtained by removing  $s-k$ rows and $t-k$ columns from $\mtrx A$.
%
%The determinant of a $k \times k$ matrix a $k \times k$ \emph{minor} of $\mtrx A$ in $\mathcal R$.
The largest integer $r$ such that there is a nonzero $r \times r$ minor of $\mtrx A$ is called the \emph{rank} of $\mtrx A$ in $\mathcal R$ and denoted by $\rank_{\mathcal R} \mtrx A$. A matrix $\mtrx A \in \Ring^{s \times s}$ is invertible in $\Ring$ if and only if the determinant of $\mtrx A$ is invertible as an element of $\Ring$ (cf. \cite[Thm.~2.1]{basicalgebraI}).

For $\mtrx A \in \Ints^{s \times t}$, we denote the gcd of all $k \times k$ minors of $\mtrx A$ by $g_k(\mtrx A)$. If $\delta \in \Ints$ is some minor of $\mtrx A$, the corresponding minor of $\mtrx A\toZm$ is $\delta\toZm$. Hence, $\rank_{\Ints_m} \mtrx A = \rank_{\Ints} \mtrx A$ for all $m \nmid g_r(\mtrx A)$, where $r= \rank_{\Ints} \mtrx A$.\footnote{In particular, the requirement $a \nmid b$ is satisfied if $a > b$.} Also,
\begin{equation}
	\rank_{\GF_q} \mtrx A = \rank_{\Ints} \mtrx A \quad \Leftrightarrow \quad p \nmid g_r(\mtrx A). \label{eq:rank_Fq_iff_p_nmid}
\end{equation}
It is known \cite[Cor.~1.13, Cor.~1.20]{normanalgebra} that there exists a unique diagonal matrix $\mtrx D = \diag(d_1, \dotsc, d_{\min(s, t)}) \in \Ints^{s \times t}$ called the \emph{Smith normal form} of $\mtrx A$, with the following properties.
\begin{enumerate}
	\item $\mtrx D = \mtrx P \mtrx A \mtrx Q$ for some matrices $\mtrx P \in \Ints^{s \times s}$ and $\mtrx Q \in \Ints^{t \times t}$ invertible in $\Ints$,
	\item $d_{i} \mid d_{i+1}$ for $i \in [\min(s,t)-1]$,
	\item $d_1 d_2 \dotsb d_i = g_i(\mtrx A)$ for $i \in [\min(s,t)]$.
\end{enumerate}

The diagonal elements $d_1, \dotsc, d_{\min(s, t)}$ are \emph{invariant factors}, and $d_i = 0$ if and only if $i > \rank_{\Ints} \mtrx A$. While $\mtrx D$ is unique, the matrices $\mtrx P$ and $\mtrx Q$ are not unique in the general case. It is also important to mention that the Smith normal form is defined for matrices over \emph{principal ideal domains (PIDs)}. For example, $\Ints$ is a PID while $\Ints_m$ is not (in general).

Random variables are labeled by capital roman letters and we write $X \sim Y$ to indicate that $X$ and $Y$ are identically distributed. Moreover, $X \sim \unif(\mathcal S)$ means that $X$ is uniformly distributed over the set $\mathcal S$. We use $\log$ to denote logarithm base-$2$, although most statements hold for an arbitrary constant base. We denote the entropy in bits and $q$-ary units by $\HH(\cdot)$ and $\HH_q(\cdot)$, respectively, and $\II(\cdot ; \cdot)$ denotes mutual information. The binary entropy function is denoted by $\hh(\cdot)$.

The notation $O(\phi(x))$ stands for any function $\psi(x)$ in $x$ such that $| \psi(x) / \phi(x)| < \const B$ for all large enough $x$ and some constant $\const B > 0$ independent of $x$. Also, $o(\phi(x))$ represents any $\psi(x)$ such that $\lim_{x \to \infty} \psi(x) / \phi(x) = 0$. In particular, $O(1)$ is any bounded function and $o(1)$ is any function that converges to zero as $x \to \infty$.

\subsection{Private Computation}
Suppose we have $n$ noncommunicating databases, each storing duplicated data: $f$ messages subpacketized into $\lambda$ parts, each part denoted as $X_i^{(j)} \in \GF_q$ for $i \in [f]$ and $j \in [\lambda]$.
%\footnote{In general, usually an arbitrary finite field is assumed in the literature. However, in this work, we restrict to prime fields, i.e., $q$ is assumed to be a prime number.} 
The subpackets are considered mutually independent and uniformly drawn from $\GF_q$. There are $\mu$ public functions $\varphi_1, %\varphi_2, 
\dotsc, \varphi_\mu$, where $\varphi_i : \GF_q^f \to \GF_q$ for $i \in [\mu]$. 
%The functions are known both to all the databases and the user. 
The user randomly chooses a secret index $V \sim \unif([\mu])$ and wants to retrieve %$\varphi_V$ calculated on all messages as follows:
\begin{align*}
\vec F_V = \left(\varphi_V(\vec{X}^{(1)}),%\varphi_V(\vec{X}^{(2)}),
\dotsc,\varphi_V(\vec{X}^{(\lambda)})\right) \in \GF_q^\lambda,
\end{align*}
where $\vec{X}^{(j)} \triangleq (X_1^{(j)},\ldots,X_f^{(j)})$, $j\in[\lambda]$, without revealing any information about $V$. To achieve that, the user and the databases employ the following scheme.
\begin{enumerate}
	\item The user generates secret randomness $R$, computes queries $Q_j = Q_j(V, R)$, $j \in [n]$, and sends the $j$-th query to the $j$-th database.
	\item Based on $Q_j$ and all the messages, the $j$-th database computes the response $A_j = A_j\left(Q_j, \vec{X}^{(1)}, \ldots,\vec{X}^{(\lambda)}\right)$ and sends it back to the user.
	\item Using all available information, the user can recover $\vec F_V$.
\end{enumerate}

Formally, we require the scheme to satisfy
\begin{center}
	\begin{tabular}{ll}
		Privacy:  & $\II(V ; Q_j) = 0$, for all $j \in [n]$, \\
		Recovery: & $\HH(\vec F_V \mid V, R, A_1, \dotsc, A_n) = 0$.\\
	\end{tabular}
\end{center}

%Since each $\mathcal Q_i$ is a deterministic function of $V$ and $R$, the latter can be simplified:
%\begin{gather*}
%\intertext{[Recovery]}
%\HH(\vec F_V \mid V, R, \mathcal A_1, \dotsc, \mathcal A_n) = 0.
%\end{gather*}

\begin{define}
The \emph{download rate} of a PC scheme over the field $\GF_q$, referred to as the PC rate, is defined as
\begin{multline*}
\const{R} = \const R(n, f, \mu,  \{\varphi_i\}, \lambda, \{Q_j\}, \{A_j\}, q) 
          \triangleq \frac{\min_{v \in [\mu]} \HH(\vec F_v)}{\Delta},
\end{multline*}
where $\Delta$ is the expected total number of downloaded bits, referred to as the \emph{download cost}.
%\footnote{Note that $\Delta$ should be the same for any $v \in [\mu]$,  as otherwise the databases will be able to obtain at least some information about the secret index $V$ from the sizes of their responses. % \todo{Verify!}}
The supremum of all achievable rates for all choices of $\lambda$, $\{Q_j\}$, and $\{A_j\}$
%field sizes $q$ 
is the \emph{PC capacity} over $\GF_q$, $\const C_{\mathrm{PC}}(n, f, \mu, \{\varphi_i\}, q)$.
\end{define}

In case $\mu = f$ and $\varphi_i(x_1, \dotsc, x_f) = x_i$ for $i \in [f]$, PC reduces to PIR with capacity
$\const{C}_{\mathrm{PIR}}(n, f) \triangleq ( 1 + 1/n + 1/n^2 + \dotsb + 1/n^{f-1})^{-1}$  \cite{SunJafar17_1}. Note that $\const{C}_{\mathrm{PIR}}$ is independent of $q$.
%\[
%\const{C}_{\mathrm{PIR}}(n, f) \triangleq \left( 1 + \frac 1n + \frac{1}{n^2} + \dotsb + \frac{1}{n^{f-1}}\right)^{-1}.
%\]

The case when $\varphi_1, %\varphi_2, 
\dotsc, \varphi_\mu$ are linear functions described by a matrix of coefficients $\mtrx A \in \GF_q^{\mu \times f}$ without zero rows, is referred to as private linear computation (PLC). Its capacity $\const{C}_\mathrm{PLC}$ only depends on $n$ and $r = \rank_{\GF_q}\mtrx A$, and it holds that $\const{C}_{\mathrm{PLC}}(n,r) = \const{C}_{\mathrm{PIR}}(n, r)$ \cite{SunJafar19_2}.\footnote{In \cite{SunJafar19_2}, the authors assume  the messages are among the functions, e.g., $\varphi_i(x_1, \dotsc, x_f) = x_i$ for $i \in [f]$. However, this is not required as we can define linearly independent functions as new variables and express other functions in these variables.}

%The straightforward approach to the private computation would be to define new $Y_i^{(j)} = \varphi_i(X_1^{(j)}, \dotsc, X_f^{(j)})$, $i \in [\mu]$ and $j \in [\lambda]$, and run any of the PIR schemes with $n$ databases and $\mu$ messages oblivious to dependencies between $Y_i^{(j)}$.

In this work, we consider private monomial computation (PMC), i.e., the case when $\varphi_i(x_1, \dotsc, x_f) = x_1^{a_{i1}} x_2^{a_{i2}} \dotsb x_f^{a_{if}}$, $i \in [\mu]$, where $a_{ij}\in\Ints$. % are \jjdel{nonnegative} integers.
The monomials can be  described by a matrix of degrees $\mtrx A = (a_{ij}) \in \Ints^{\mu \times f}$, and we assume there are no constant functions, i.e.,  no zero rows in $\mtrx A$. The capacity of PMC is denoted by $\const{C}_{\mathrm{PMC}}(n, f, \mu, \mtrx A, q)$.
%, depends in general on both $\mu$, $\mtrx A$ and $q$.

\section{Entropies of Linear Functions and Monomials}\label{sec:entropies}
\begin{lemma}\label{lem:H-ay}
	Let $a \in \Ints$ and $Y \sim \unif(\Ints_m)$. Then,
	\[
%	\HH(a Y) = \log \frac{m}{\gcd(a, m)}.
	\HH(a Y) = \HH(a\toZm Y) = \log m - \log \gcd (a, m).
	\]
\end{lemma}
\begin{IEEEproof}
	From the theory of linear congruences \cite[Sec.~5, Thm.~1]{elemnumtheory}, the equation $a y = b$ has $d = \gcd(a, m)$ solutions in $\Ints_m$ if $d \mid b$ and no solutions otherwise. Therefore, the random variable $aY$ takes $m / d$ different values from $\Ints_m$ equiprobably, and the required statement follows.
\end{IEEEproof}

\begin{lemma}\label{lem:H-Ay}
	Let $\mathsf A \in \Ints^{s \times t}$ be a fixed matrix whose invariant factors are $d_1, \dotsc, d_{\min(s,t)}$. Let $\vec Y = (Y_1, \dotsc, Y_t) \sim \unif(\Ints_m^t)$, $r = \rank_{\Ints} \mtrx A$, and $r' = \rank_{\Ints_m} \mtrx A\toZm$. Then, 
	\begin{align}
	\HH(\mathsf A \vec Y) &= r\log m - \sum_{i=1}^{r} \log \gcd(d_i, m) \label{eq:something}\\
	                      &= r'\log m - \sum_{i=1}^{r'} \log \gcd(d_i, m).\label{eq:because-of-divisions}
	\end{align}
%	where $\vec Y = (Y_1, \dotsc, Y_t)\transp$,  .
\end{lemma}
\begin{IEEEproof}
	Recall that, since $\vec Y$ is defined over $\Ints_m^t$, the operations in $\mtrx A \vec Y$ are  over $\Ints_m$. In other words, $\mtrx A \vec Y$ is a shorthand for $\mtrx A\toZm \vec Y$.
	
	Let $\mtrx D = \mtrx P \mtrx A \mtrx Q$ be the Smith normal form of $\mtrx A$, where both $\mtrx P \in \Ints^{s \times s}$ and $\mtrx Q \in \Ints^{t \times t}$ are invertible over $\Ints$ (i.e., their determinants are $\pm 1$) and $\mtrx D = \diag(d_1, \dotsc, d_r, 0, \dotsc, 0)$. After taking modulo $m$ from both sides, we obtain $\mtrx D\toZm = \mtrx P\toZm \mtrx A\toZm \mtrx Q\toZm$, where $\mtrx P\toZm$ and $\mtrx Q\toZm$ are both invertible over $\Ints_m$ (their determinants are $\pm 1$ in $\Ints_m$ too) and $\mtrx D\toZm = \diag(d_1\toZm, \dotsc, d_r\toZm, 0, \dotsc, 0)$. 
%	Next, $d_i\toZm \neq 0$, for $i \le r'$, and $d_i\toZm = 0$, for $i > r'$. 
	Therefore,
	\begin{align*}
	&\HH(\mtrx D\toZm \vec Y) = \HH(\mtrx P\toZm (\mtrx A\toZm \mtrx Q\toZm \vec Y)) = \HH(\mtrx A\toZm \mtrx Q\toZm \vec Y) \\ %\label{eq:because-of-P} \\
	&\quad= \HH(\mtrx A\toZm (\mtrx Q\toZm \vec Y)) = \HH(\mtrx A\toZm \vec Y) = \HH(\mtrx A \vec Y), %\label{eq:because-of-Q}
	\end{align*}
	because $\mtrx P\toZm$ and $\mtrx Q\toZm$ are invertible over $\Ints_m$, and multiplication from the left by an invertible matrix is a bijection. Thus, we can consider $\HH(\mtrx D\toZm \vec Y)$ instead of $\HH(\mtrx A \vec Y)$.
	But $\mtrx D\toZm \vec Y = (d_1\toZm Y_1, \dotsc, d_r\toZm Y_r, 0, \dotsc, 0)$ with mutually independent entries. Hence,
	\begin{align*}
	\HH(\mtrx D\toZm \vec Y) &= \sum_{i=1}^r \HH(d_i\toZm Y_i) \\
	&\stackrel{\text{Lem.~\ref{lem:H-ay}}}= r \log m - \sum_{i=1}^r \log \gcd(d_i, m).
	\end{align*}
	Finally, \cref{eq:because-of-divisions} holds because $m \mid d_i$ for $i > r'$ and hence $\gcd(d_i,m)=m$.
\end{IEEEproof}

\begin{cor}\label{cor:H-Ay-asym}
	In the setting of \cref{lem:H-Ay}, 
	$\HH(\mtrx A \vec Y)= r\log m + O(1),\text{ as }m \to \infty$, where $r = \rank_{\Ints} \mtrx A$.
\end{cor}
\begin{IEEEproof}
	For all $m > d_r$ and all $i \in [\min(s,t)]$, it holds that $d_i\toZm = d_i$. In this case, $r' = r$ and
	\begin{equation}\label{eq:HAY-lb}
\begin{aligned}
\HH(\mtrx A \vec Y) &= r \log m - \sum_{i=1}^r \log \gcd(d_i, m) \\
&\ge r \log m - \log \prod_{i=1}^r d_i = r \log m - \log g_r(\mtrx A).
\end{aligned}
\end{equation}
	On the other hand,
	\begin{align}\label{eq:HAY-ub}
\HH(\mtrx A \vec Y) &= r \log m - \sum_{i=1}^r \log \gcd(d_i, m) \le r \log m.
\end{align}
	
	We note that both \cref{eq:HAY-lb} and \cref{eq:HAY-ub}
% note: do not put the references about in one \cref{} as they are further referred in text in exactly the order above
	are attained for infinitely many values of $m$, e.g., for $m=u g_r(\mtrx A)$ and $m = 1 + u g_r(\mtrx A)$, respectively (for any positive integer $u$). In other words, $\HH(\mtrx A \vec Y)$ does not converge as $m \to \infty$.
	
	Finally, as $\log g_r(\mtrx A)$ does not depend on $m$, we have
	\[
	\HH(\mtrx A \vec Y)= r \log m + O(1), \text{ as }m\to\infty. \IEEEQEDhereeqn
	\]
\end{IEEEproof}

Next, we present some results on entropies of monomials over finite fields. The key idea is to use the bijection of $\dlog$ and treat a special case of zero separately.

\begin{lemma}\label{lem:H-one-mono}
	Let $a_1, \dotsc, a_t \in \Ints$, $X_1, \dotsc, X_t \sim \unif(\GF_q)$ be mutually independent, $\tau$ be the number of nonzeros among $a_1, \dotsc, a_t$, and $\pi = \left( 1 - 1/q \right)^\tau$. Then, %In this case, 
	\begin{align*}
	\HH(X_1^{a_1}X_2^{a_2} \dotsb X_t^{a_t}) = \hh(\pi) + \pi \log \frac{q-1}{\gcd(a_1, \dotsc, a_t, q-1)}.
	\end{align*}
	Moreover, if not all $a_1, \dotsc, a_t$ are zeros,
	\begin{align*}
	\HH_q(X_1^{a_1}X_2^{a_2} \dotsb X_t^{a_t}) \xrightarrow[q \to \infty]{} 1.
	\end{align*}
\end{lemma}
\begin{IEEEproof}
	%We understand $a_i = 0$ as the variable $X_i$ is not present in the monomial.
	If $a_i = 0$, the variable $X_i$ is not present in the monomial.
	Hence, we can exclude such variables and assume $a_1, \dotsc, a_\tau \in \Ints \setminus \{0\}$. Dropping zero arguments of the gcd above does not change its value either.
	
	Let $M = X_1^{a_1}X_2^{a_2} \dotsb X_\tau^{a_\tau}$. Define $Z = 0$ if $M=0$ and $Z=1$ otherwise.
%		\[
%	Z = \begin{cases}
%	0, & \text{if $M=0$},\\
%	1, & \text{otherwise}.
%	\end{cases}
%	\]
	Then, $\pi = \Prob{M \neq 0} = \Prob{Z = 1}$ and
	\begin{align*}
	\HH(M) &= \HH(Z) + \HH(M \mid Z) - \HH(Z \mid M) \\
	&= \hh(\pi) + \HH(M \mid Z = 0) (1-\pi) + \HH(M \mid Z = 1) \pi \\
	&= \hh(\pi) + \pi \HH(M \mid M \neq 0).
	\end{align*} 
	
	Now, $M \neq 0$ if and only if none of $X_1, \dotsc, X_\tau$ is zero. In this case, all $X_1, \dotsc, X_\tau \in \GF_q^*$ and we can define $Y_j = \dlog X_j \in \Ints_{q-1}$ for $j \in [\tau]$ and $L' = \dlog M = a_1 Y_1 + \dotsb + a_\tau Y_\tau \in \Ints_{q-1}$. Since $\dlog$ is bijective, $Y_1, \dotsc, Y_\tau \sim \unif(\Ints_{q-1})$ and $\HH(M \mid M \neq 0) = \HH(L')$. 
	By applying \cref{lem:H-Ay} with $m=q-1$, $s = 1$, $r=1$, and $d_1 = \gcd(a_1, \dotsc, a_\tau)$, we get
%	. If not all $a_1, \dotsc, a_\tau$ are divisible by $q-1$, we have $r' = 1$ and
	\begin{align*}
	\HH(L') &= \log \frac{q-1}{\gcd(a_1, \dotsc, a_\tau, q-1)}.
	\end{align*}
%	Otherwise, if all $a_1, \dotsc, a_\tau$ are divisible by $q-1$, we have $r'=0$ and $\HH(L') = 0$. On the other hand, $\gcd(a_1, \dotsc, a_\tau, q-1) = q-1$, and it follows that the formula above holds in this case as well.
	
%<<<<<<< HEAD
	Further,
	% for all $q > \max(|a_1|, \dotsc, |a_t|)$, $t'$ is equal to the number of non-zeros among $a_1, \dotsc, a_t$ and, therefore, is constant. Consequently, 
	as $q \to \infty$, $\pi \to 1$ and therefore $\hh(\pi) \to 0$. Additionally, %if at least one of $a_1, \dotsc, a_t$ is non-zero, 
	$\gcd(a_1, \dotsc, a_\tau, q-1) \le \min(|a_1|, \dotsc, |a_\tau|) = O(1)$, as $q \to \infty$.
%=======
%	Further, as $q \to \infty$, we have $\pi \to 1$ and therefore $\hh(\pi) \to 0$. Additionally, if at least some of $a_1, \dotsc, a_t$ are non-zero, $\gcd(a_1, \dotsc, a_t, q-1) \le \max(|a_1|, \dotsc, |a_t|) = O(1)$, as $q \to \infty$.
%>>>>>>> 2a2dc3ea1cb550578a43953338115826e1b4dd84
	%
	Finally,
	\[
	\HH_q(X_1^{a_1}X_2^{a_2} \dotsb X_t^{a_t}) = \frac{\HH(X_1^{a_1}X_2^{a_2} \dotsb X_t^{a_t})}{\log q} \xrightarrow[q \to \infty]{} 1. \IEEEQEDhereeqn
	\]
\end{IEEEproof}

\begin{thm}\label{thm:monomials-entropy}
	Let $\mtrx A \in \Ints^{s \times t}$ be a fixed matrix of coefficients with rank $r=\rank_{\Ints} \mtrx A$. Let $X_1, \dotsc, X_t \sim \unif(\GF_q)$ be mutually independent. For $i \in [s]$, define $M_i = X_1^{a_{i1}} X_2^{a_{i2}} \dotsb X_t^{a_{it}} \in \GF_q$ and $\vec M = (M_1, \dotsc, M_s)$. Then,
	\[
	\HH(\vec M) = r \log q + O(1), \text{ as }q\to\infty.
	\]
\end{thm}
\begin{IEEEproof}
	First, if there is a zero column in $\mtrx A$, we can drop the corresponding variable, as it does not influence either the values of any of the monomials or $\rank_{\Ints} \mtrx A$. Thus, for the remainder of the proof, we assume there are no zero columns in $\mtrx A$, and we also consider values of $q$ large enough so that there are no zero columns in $\mtrx A^{\langle q-1 \rangle}$ as well. 
	%For example, it is enough to require that $q$ is larger than any absolute value of the matrix entry.
	
	Define $Z=0$ if $X_1 X_2 \dotsb X_t = 0$ and $Z=1$ otherwise.
%	\[
%	Z = \begin{cases}
%	0, &\text{if $X_1 X_2 \dotsb X_t = 0$},\\
%	1, &\text{otherwise.}
%	\end{cases}
%	\]
	It holds that $\pi = \Prob{Z=1} = (1 - 1/q )^t$. Moreover, $Z = 0$ if and only if any of the monomials $M_1, \dotsc, M_s$ is zero. Hence, $\HH(Z \mid \vec M) = 0$ and we have
	\begin{align*}
	\HH(\vec M) &= \HH(Z) + \HH(\vec M \mid Z) - \HH(Z \mid \vec M) \\
	&= \hh(\pi) + (1-\pi) \HH(\vec M \mid Z = 0) + \pi \HH(\vec M \mid Z = 1).
	\end{align*}
	
	Next, $Z = 1$ if and only if none of $X_1, \dotsc, X_t$ is zero, i.e., all $X_1, \dotsc, X_t \in \GF_q^*$. In this case, we can define $Y_j = \dlog X_j \in \Ints_{q-1}$, for $j \in [t]$, $L_i' = \dlog M_i = a_{i1} Y_1 + \dotsb + a_{it} Y_t \in \Ints_{q-1}$, for $i \in [s]$, and $\vec L' = (L_1', \dotsc, L_s')$. Then, $\HH(\vec L') = \HH(\vec M \mid Z = 1)$ and
	\begin{align*}
	&|\HH(\vec M) - \HH(\vec L') | = |\HH(\vec M) - \HH(\vec M \mid Z=1)| \\
	&\ =|\hh(\pi) + (1-\pi)\HH(\vec M | Z = 0) + (\pi-1)\HH(\vec M \mid Z=1)| \\
	&\ \le \hh(\pi) + (1-\pi)|\HH(\vec M | Z = 0) - \HH(\vec M | Z = 1)| \\
	&\ \le \hh(\pi) + s(1-\pi) \log q = o(1),%O \left( \frac{\log q}{q} \right), 
	\text{ as } q \to \infty.
	\end{align*}
	
%	Moreover, $\hh(\pi) + s(1-\pi)\log q <\log 2 + st$.
	From \cref{cor:H-Ay-asym} with $m=q-1$, we have $\HH(\vec L') = r \log(q-1) + O(1) = r \log q + O(1)$, as $q \to \infty$. Finally, 
	\[
%	\HH(\vec M) = \HH(\vec L) + O \left( \frac{\log q}{q} \right) = r \log q + O(1). \IEEEQEDhereeqn
	\HH(\vec M) = \HH(\vec L') + o(1) = r \log q + O(1), \text{ as }q\to\infty. \IEEEQEDhereeqn
	\]
\end{IEEEproof}

\begin{cor}\label{cor:HMapproxHL}
	In the setting of \cref{thm:monomials-entropy}, consider $q=p^k$ with $p \nmid g_r(\mtrx A)$. Then,
	\[
	\left| \HH_q(\vec{M}) - \HH_q(\vec{L}) \right| = o(1), \text{ as }q\to\infty,
	\]
	where
	%\[
	$L_i = a_{i1}X_1  + \dotsb + a_{it} X_t \in \GF_q$ for $i \in [s]$, and $\vec L = (L_1, \dotsc, L_s)$.%
	\footnote{In contrast to \cref{lem:H-Ay} and \cref{cor:H-Ay-asym}, $\vec L$ is defined over the field.}
	%\]
\end{cor}
\begin{IEEEproof}
	As $\mtrx A$ defines a linear transformation of a vector space over $\GF_q$, $\HH(\vec L) = \rank_{\GF_q} \mtrx A \cdot \log q$. From \cref{eq:rank_Fq_iff_p_nmid} and since $p \nmid g_r(\mtrx A)$, we obtain $\rank_{\GF_q} \mtrx A = \rank_{\Ints} \mtrx A = r$. Next, from \cref{thm:monomials-entropy}, as $q\to\infty$, 
	\[
	\left| \HH_q(\vec M) - \HH_q(\vec L) \right| = \frac{\left| \HH(\vec M) - \HH(\vec L) \right|}{\log q} = \frac{O(1)}{\log q} = o(1).
	\IEEEQEDhereeqn
	\]
\end{IEEEproof}

Note that we do not require $p$ to be either fixed or infinitely large. However, all primes $p > g_r(\mtrx A)$ satisfy the requirement $p \nmid g_r(\mtrx A)$. \Cref{cor:HMapproxHL} states that the entropy of any fixed set of monomials is equal to the entropy of the corresponding set of linear functions (i.e., defined by the same matrix $\mtrx A$), both over $\GF_q$, when $p \nmid g_r(\mtrx A)$ and as $q$ approaches infinity. Moreover, this also holds for conditional entropies consisting of various sets of monomials because they can be expressed as a difference of two unconditional entropies. This key observation is further used in \cref{sec:PMC-scheme}.

\section{Achievable Scheme}
\subsection{Sun--Jafar Scheme for Private Linear Computation}
We build our PMC achievable scheme based on the Sun--Jafar scheme for PLC (\cite[Alg.~1]{SunJafar19_2}, referred to as \emph{PC} there). %\footnote{Despite the name, Sun--Jafar \emph{PC} scheme is for linear case only.}
Due to lack of space, we do not present their scheme in all details and refer the reader to \cite{SunJafar19_2} for a full description and analysis. Here, we briefly repeat the facts (in our notation) essential for further discussion.

The Sun--Jafar scheme uses $\lambda=n^\mu$ subpackets. From each of the $n$ databases, the user downloads symbols in $\mu$ blocks. The $b$-th block, $b \in [\mu]$, of each database consists of $(n-1)^{b-1} \binom{\mu}{b}$ symbols, and each symbol is a linear combination (using only coefficients $\pm 1$) of $b$ judiciously chosen pieces $\varphi_u(\vec{X}^{(j)})$ for different values of $u \in [\mu]$ and $j \in [\lambda]$. Since all $\varphi_u$ are linear combinations, each symbol the user downloads is some linear combination of $\{X_i^{(j)}\}$. The user's randomized queries define which linear combinations the databases will reply with. The queries enforce symmetry across databases and function evaluation symmetry within symbols downloaded from each database. This ensures privacy of the user.

A crucial observation is that $(n-1)^{b-1} \binom{\mu - r}{b}$ of the symbols in block $b$ of each database are redundant based on side information downloaded from other databases. More precisely, these redundant symbols are linear combinations of other symbols in block $b$ from the same database as well as symbols downloaded from other databases. Hence, they need not to be downloaded, as the user can reconstruct them offline. 
%This allows the authors to employ Slepian--Wolf source coding with side information \cite{slepianwolf73_1}. In other words, before sending, a database can encode the whole block $b$ into no more than $(n-1)^{b-1} \left( \binom{\mu}{b} - \binom{\mu-r}{b} \right)$ $q$-ary symbols, without knowledge of exactly what side information is actually available to the user. 
This preserves the user's privacy while reducing the download cost to the value corresponding to the PLC capacity. A distinctive property of the Sun--Jafar scheme is that it is oblivious to the  coefficients of the linear functions $\varphi_v$. It is only the number of them, $\mu$, that matters. Furthermore, the scheme can be used for PIR if $\mu = f$ and the linear functions are the messages, i.e., $\varphi_i(x_1, \dotsc, x_f) = x_i$ for $i \in [f]$. In this case, there are no redundant symbols in any block.

\subsection{Private Monomial Computation}\label{sec:PMC-scheme}
%We propose the following achievable scheme with subpacketization size $\lambda = n^\mu$.

Let $\lambda = n^\mu$ and suppose that none of $\{X_i^{(j)}\}$ equals zero. Then we can construct a \emph{multiplicative} scheme  by substituting each linear combination of $\{ \varphi_v \}$ in the Sun--Jafar scheme with a corresponding multiplicative combination. For example, if at some step the user downloads the symbol $\varphi_1(\vec{X}^{(j_1)}) + \varphi_2(\vec{X}^{(j_2)}) - \varphi_3(\vec{X}^{(j_3)})$, $j_1,j_2,j_3\in[\lambda]$, then the corresponding multiplicative combination is $\varphi_1(\vec{X}^{(j_1)}) \varphi_2(\vec{X}^{(j_2)}) \bigl(\varphi_3(\vec{X}^{(j_3)})\bigr)^{-1}$, 
% \end{align*}
where the functions $\varphi_v$ now denote the corresponding monomials. Since there are no zeros among $\{X_i^{(j)}\}$, all operations are valid and ensure correct reconstruction of the monomial of interest. Moreover, from \cref{cor:HMapproxHL}, when $p \nmid g_r(\mtrx A)$ and  as $q \to \infty$, the entropies of all the symbols as well as the entropy of each block $b$ conditioned on the side information received from other databases converge to those of the Sun--Jafar scheme. This means that in the multiplicative scheme above, a database can also encode the whole $b$-th block into no more than $(n-1)^{b-1}\left( \binom{\mu}{b} - \binom{\mu-r}{b} \right)$ $q$-ary symbols,  resulting in the same download cost as in the Sun--Jafar scheme. Since there is only a finite number of entropies involved, we can satisfy the requirement on $p$ from \cref{cor:HMapproxHL} for all of them simultaneously, e.g., by requiring $p$ to be large enough (but not necessarily approaching infinity).

Now, in case any of $\{X_i^{(j)}\}$ equals zero, we can ignore dependencies between the monomials and run a PIR scheme, for example, the same Sun--Jafar scheme in PIR mode for $\mu$ messages. Altogether, our scheme is as follows.

%\vspace{-1ex}
\begin{algorithm}
	\caption{PMC Scheme}
	\label{alg:PMC-scheme}
	
%	\begin{algorithmic}[1]
%		Send to each database queries both for MS and PIR scheme. Also, ask one database whether $0 \in \{X_i^{(j)}\}$.\label{alg:PMC-scheme:communicate-bit}
	
		\eIf{there are no zeros among $\{ X_i^{(j)}\}$ and $\mu > r$}{	
			Each database replies according to the multiplicative scheme.\label{alg:PMC-scheme:mult}
		}{
			Each database replies according to the Sun--Jafar scheme in PIR mode oblivious to the dependencies between the monomials.\label{alg:PMC-scheme:PIR}
		}
%	\end{algorithmic}
\end{algorithm}
%\vspace{-1ex}

Note that the queries of both schemes need to be uploaded since the user does not know  if there are zeros among $\{ X_i^{(j)}\}$. Moreover, the user can  determine which scheme is used  (\cref{alg:PMC-scheme:mult} or \cref{alg:PMC-scheme:PIR}) from $(r,\mu)$ and the size of the responses (the size is smaller for the multiplicative scheme provided $r < \mu$). %In the case $\mu = r$, the databases should always choose the scenario in \cref{alg:PMC-scheme:PIR}.}

We note that privacy of the user in the suggested PMC scheme is inherited from the privacy of the Sun--Jafar scheme.

%\begin{thm}\label{thm:PMC-achievable-rate}
%	For PMC with $n$ databases, $f$ messages, and $\mu$ monomials defined by a degree matrix $\mtrx A \in \Ints^{\mu \times f}$ of rank $r = \rank_{\Ints} \mtrx A$, the rate $\const R = \const C_{\mathrm{PIR}}(n,r)$ is achievable for large enough $p$ as $q \to \infty$.
%\end{thm}
%\begin{IEEEproof}
%	
%\end{IEEEproof}

\begin{thm}\label{thm:PMC-asym-cap}
	For PMC with $n$ databases, $f$ messages, and $\mu$ monomials defined by a degree matrix $\mtrx A \in \Ints^{\mu \times f}$ of rank $r = \rank_{\Ints} \mtrx A$, for $p \nmid g_r(\mtrx A)$ and as $q \to \infty$, the PMC capacity converges to that of PIR: $\const C_{\mathrm{PMC}}(n, f, \mu, \mtrx A, q) \to \const C_{\mathrm{PIR}}(n,r)$.
\end{thm}
\begin{IEEEproof}
	First, we show that the PC rate $\const C_{\mathrm{PIR}}(n,r)$ is achievable by \cref{alg:PMC-scheme}. %At \cref{alg:PMC-scheme:communicate-bit}, we download $1$ bit, i.e., $\log_q 2$ $q$-ary units. 
	For \cref{alg:PMC-scheme:mult}, for  $p \nmid g_r(\mtrx A)$ and as $q \to \infty$, the download cost measured in $q$-ary units converges to $n^\mu / \const{C}_{\mathrm{PLC}}(n,r) = n^\mu / \const{C}_{\mathrm{PIR}}(n,r)$. The download cost at \cref{alg:PMC-scheme:PIR} is $n^\mu / \const{C}_{\mathrm{PIR}}(n,\mu)$.% Finally, $1$ bit gives $\log_q 2$.
	
	The probability that none of $\{X_i^{(j)}\}$ equals zero is $\pi = (1 - 1/q )^{n^\mu f} \to 1$, as $q \to \infty$.
	%	 \[
	%	 \pi = \left( 1 - \frac 1q \right)^{n^\mu f} \xrightarrow[q \to \infty]{} 1.
	%	 \]
	Therefore, the average download cost of \cref{alg:PMC-scheme} becomes% , as $q \to \infty$,
	\begin{align*}
	%\log_q 2 + 
	n^\mu \left( \frac{\pi}{\const{C}_{\mathrm{PIR}}(n,r)} + \frac{1 - \pi}{\const{C}_{\mathrm{PIR}}(n,\mu)} \right)\xrightarrow[q \to \infty]{} 
	\frac{n^\mu}{\const{C}_{\mathrm{PIR}}(n,r)}.
	\end{align*}
	On the other hand, from \cref{lem:H-one-mono}, it follows that %\footnote{We do not write the subpacket index for $\vec X$ because it does not change the entropy.} 
	\begin{align*}
	\min_{v \in [\mu]} \HH_q(\vec F_v) = n^\mu \cdot \min_{v \in [\mu]} \HH_q(\varphi_v(\vec X^{(1)})) 
	\xrightarrow[q \to \infty]{} n^\mu.
	\end{align*}
	Altogether, we have that the download rate of our PMC scheme converges to the PIR capacity for $r$ messages. 
	
	It remains to prove the converse, i.e., showing that $\const{C}_{\mathrm{PIR}}(n,r)$ is an upper (or outer) bound on the PC capacity.
	%
	%
	%an upper bound on the capacity. 
	%
	%
	For that, we consider the general converse in \cite[Thm.~1]{ObeadLinRosnesKliewer19_2} and show that, for $q \to \infty$ and provided $p \nmid g_r(\mtrx A)$, the upper bounds from  \cite[Thm.~1]{ObeadLinRosnesKliewer19_2}  coincide for the monomial and linear cases with the same matrix $\mtrx A$. Note that \cite[Thm.~1]{ObeadLinRosnesKliewer19_2} gives $\mu!$ upper bounds on the PC capacity  (according to the number of permutations of $\mu$ functions). For the linear case, the outer bounds in \cite[Thm.~1]{ObeadLinRosnesKliewer19_2} reduce to $\const C_{\mathrm{PIR}}(n,r)$, independent of $q$. 
	In general, for a fixed permutation, the bound depends on $\min_{v \in [\mu]} \HH_q(\varphi_v(\vec X^{(1)}))$ and joint entropies of different subsets of function evaluations. Then, it follows from the key observation in \cref{sec:entropies} that this bound is coinciding for the monomial and linear cases as $q \to \infty$, provided $p \nmid g_r(\mtrx A)$ (details omitted for brevity).
\end{IEEEproof}

\begin{cor} \label{cor:scheme-is-cap-ach}
	In the setting of \cref{thm:PMC-asym-cap}, the scheme in \cref{alg:PMC-scheme} is capacity-achieving for $p \nmid g_r(\mtrx A)$ and as $q \to \infty$. % for $q \to \infty$.
%	For PMC with $n$ databases, $f$ messages, and $\mu \ge f$ monomials defined by a degree matrix $\mtrx A \in \Ints^{\mu \times f}$, the PMC scheme of \cref{alg:PMC-scheme} is capacity-achieving, for large enough $\characteristic \GF_q$ and as $q \to \infty$,  when $\varphi_i(x_1, \dotsc, x_f) = x_i$ for $i \in [f]$. %Then the scheme in \cref{alg:PMC-scheme} achieves the capacity of this problem.
\end{cor}
%\begin{IEEEproof}
%Since $\varphi_i(x_1, \dotsc, x_f) = x_i$ for $i \in [f]$, $r = \rank_{\Ints} \mtrx A = f$ and the PMC rate of \cref{alg:PMC-scheme}, for large enough $\characteristic \GF_q$ and as $q \to \infty$, is $\const R = \const C_{\mathrm{PIR}}(n,f)$. Moreover, in this special case,  PMC is an extension of PIR. Hence, $\const{C}_{\mathrm{PMC}}(n, f, \mu, \mtrx A, q) \leq \const{C}_{\mathrm{PIR}}(n,f)$, and the result follows.
%\end{IEEEproof}
%
%Note that we are only able to prove that the rate from \cref{thm:PMC-achievable-rate} is equal to the PMC capacity $\const{C}_{\mathrm{PMC}}(n, f, \mu, \mtrx A, q)$ in the special setting of \cref{cor:scheme-is-cap-ach}, since the converse bound from \cite[Thm.~1]{ObeadLinRosnesKliewer19_2} (which is the  best know outer bound) can be higher than  $\const{C}_{\mathrm{PIR}}(n,r)$ for a finite $q$.

Note that we prove that the scheme in \cref{alg:PMC-scheme} is capacity-achieving only for asymptotic $q$ and provided  $p \nmid g_r(\mtrx A)$. As an example, take $\mu=f=2$, $n=2$, $\varphi_1(x_1,x_2) = x_1^2 x_2$, and $\varphi_2(x_1,x_2)=x_1 x_2^2$. Then the asymptotic PC  rate of \cref{cor:scheme-is-cap-ach}  is $\const{C}_{\mathrm{PIR}}(2,2) = 2/3$, since $r = \rank_{\Ints} \mtrx A = 2$. On the other hand, the PC capacity $\const C_{\mathrm{PC}}$ for two arbitrary functions for any finite field is known \cite[Sec.~VII, Eq.~(82)]{SunJafar19_2}. For this example, $\const C_{\mathrm{PC}} = 2 \HH / (\HH(X_1^2 X_2, X_1 X_2^2) + \HH)$, where $ \HH \triangleq \HH(X_1^2 X_2) = \HH(X_1 X_2^2)$ and the superscripts on the $X$'s have been suppressed for brevity. Finally, \cref{alg:PMC-scheme}  defaults to PIR mode and achieves the PC rate  ${2\HH}/3$, which can be shown to be  smaller than $\const C_{\mathrm{PC}}$ for any finite $q$.% Thus, for finite $q$, \cref{alg:PMC-scheme} is not capacity-achieving.}

%the general setting is still open.

\balance

\section{Conclusion}
%We derived the capacity and presented a capacity-achieving scheme for PMC for replicated non-colluding databases for the case when the candidate set of monomials contains all stored messages as a subset, by considering the case of an arbitrary large field. Furthermore, we presented formulas for the entropy of a multivariate monomial and for a set of monomials in uniformly distributed random variables over a finite field.
We derived the PMC capacity for replicated noncolluding databases, by considering the case of an arbitrary large field and under a technical condition on the size $p$ of the base field, which is satisfied, e.g., for $p$ large enough. A PMC scheme that is capacity-achieving in the above asymptotic case was also outlined.
%
%
%the case when the field size $q=p^k$ 
%
%for the case when the candidate set of monomials contains all stored messages as a subset, by considering the case of an arbitrary large field.
Furthermore, we presented formulas for the entropy of a multivariate monomial and for a set of monomials in uniformly distributed random variables over a finite field. 

%In this work, we consider private monomial computation (PMC) for replicated non-colluding databases. In PMC, a user wishes to privately retrieve an arbitrary multivariate monomial from a candidate set of monomials in $f$ messages over a finite field $\GF_q$, where $q=p^k$ is a power of a prime $p$ and $k \ge 1$,  replicated over $n$ databases. We derive the PMC capacity under a technical condition on $p$, which is satisfied, e.g.,  for $p$ large enough, and for asymptotically large $q$. Also, we present a novel PMC scheme for arbitrary $q$ that is capacity-achieving in the asymptotic case above.

\section*{Acknowledgment}
The authors would like to thank Srimathi Varadharajan and Alessandro Melloni for useful discussions.

%%%%%%%%%%%%%%%%%%%%%%%%%%%%%%%%%%%%%%%%%%%%%%%%%%%%%%%%%%%%%%%%%%%%%%%%%%%%%%%%

\bibliographystyle{IEEEtran}
\bibliography{./refs,./defshort1,./biblioHY}

\begin{thebibliography}{10}
\providecommand{\url}[1]{#1}
\csname url@samestyle\endcsname
\providecommand{\newblock}{\relax}
\providecommand{\bibinfo}[2]{#2}
\providecommand{\BIBentrySTDinterwordspacing}{\spaceskip=0pt\relax}
\providecommand{\BIBentryALTinterwordstretchfactor}{4}
\providecommand{\BIBentryALTinterwordspacing}{\spaceskip=\fontdimen2\font plus
\BIBentryALTinterwordstretchfactor\fontdimen3\font minus
  \fontdimen4\font\relax}
\providecommand{\BIBforeignlanguage}[2]{{%
\expandafter\ifx\csname l@#1\endcsname\relax
\typeout{** WARNING: IEEEtran.bst: No hyphenation pattern has been}%
\typeout{** loaded for the language `#1'. Using the pattern for}%
\typeout{** the default language instead.}%
\else
\language=\csname l@#1\endcsname
\fi
#2}}
\providecommand{\BIBdecl}{\relax}
\BIBdecl

\bibitem{SunJafar19_2}
H.~Sun and S.~A. Jafar, ``The capacity of private computation,'' \emph{IEEE
  Trans. Inf. Theory}, vol.~65, no.~6, pp. 3880--3897, Jun. 2019.

\bibitem{MirmohseniMaddahAli18_1}
M.~Mirmohseni and M.~A. Maddah-Ali, ``Private function retrieval,'' in
  \emph{Proc. Iran Workshop Commun. Inf. Theory (IWCIT)}, Tehran, Iran, Apr.
  25--26, 2018, pp. 1--6.

\bibitem{ObeadKliewer18_1}
S.~A. Obead and J.~Kliewer, ``Achievable rate of private function retrieval
  from {MDS} coded databases,'' in \emph{Proc. IEEE Int. Symp. Inf. Theory
  (ISIT)}, Vail, CO, USA, Jun. 17--22, 2018, pp. 2117--2121.

\bibitem{ObeadLinRosnesKliewer18_1}
S.~A. Obead, H.-Y. Lin, E.~Rosnes, and J.~Kliewer, ``Capacity of private linear
  computation for coded databases,'' in \emph{Proc. 56th Allerton Conf.
  Commun., Control, Comput.}, Monticello, IL, USA, Oct. 2--5, 2018, pp.
  813--820.

\bibitem{ObeadLinRosnesKliewer19_1}
------, ``Private polynomial computation for noncolluding coded databases,'' in
  \emph{Proc. IEEE Int. Symp. Inf. Theory (ISIT)}, Paris, France, Jul. 7--12,
  2019, pp. 1677--1681.

\bibitem{ObeadLinRosnesKliewer19_2}
------, ``On the capacity of private nonlinear computation for replicated
  databases,'' in \emph{Proc. IEEE Inf. Theory Workshop (ITW)}, Visby, Sweden,
  Aug. 25--28, 2019, pp. 1--5.

\bibitem{Karpuk18_1}
D.~Karpuk, ``Private computation of systematically encoded data with colluding
  servers,'' in \emph{Proc. IEEE Int. Symp. Inf. Theory (ISIT)}, Vail, CO, USA,
  Jun. 17--22, 2018, pp. 2112--2116.

\bibitem{RavivKarpuk19_1}
N.~Raviv and D.~A. Karpuk, ``Private polynomial computation from {L}agrange
  encoding,'' in \emph{Proc. IEEE Int. Symp. Inf. Theory (ISIT)}, Paris,
  France, Jul. 7--12, 2019, pp. 1672--1676.

\bibitem{basicalgebraI}
N.~Jacobson, \emph{Basic Algebra I}, 2nd~ed.\hskip 1em plus 0.5em minus
  0.4em\relax Freeman and Company, 1985.

\bibitem{normanalgebra}
C.~Norman, \emph{Finitely Generated Abelian Groups and Similarity of Matrices
  over a Field}.\hskip 1em plus 0.5em minus 0.4em\relax Springer Science \&
  Business Media, 2012.

\bibitem{SunJafar17_1}
H.~Sun and S.~A. Jafar, ``The capacity of private information retrieval,''
  \emph{IEEE Trans. Inf. Theory}, vol.~63, no.~7, pp. 4075--4088, Jul. 2017.

\bibitem{elemnumtheory}
U.~Dudley, \emph{Elementary Number Theory}, 2nd~ed.\hskip 1em plus 0.5em minus
  0.4em\relax Freeman and Company, 1978.

\end{thebibliography}

\end{document}